\documentclass[
aps,
final,
notitlepage,
oneside,
onecolumn,
nobibnotes,
nofootinbib,
superscriptaddress,
showpacs]
{revtex4}
\usepackage{graphicx}
\usepackage{amsmath}
\usepackage{amsfonts}
\usepackage{amssymb}
\usepackage{bbm}  
\usepackage{yfonts}  
\newif\ifTwoCol
\TwoColtrue  

\newcommand{\PLu}{\protect{Pl{\"u}cker}}
\newcommand{\PLL}{\protect{Pl{\"u}cker }}

\begin{document}
\title{On Spin II}
\author{\firstname{Rolf}~\surname{Dahm}}
\affiliation{Permanent address: beratung f{\"{u}}r Informationssysteme und Systemintegration, G{\"{a}}rtnergasse 1, D-55116 Mainz, Germany}

\begin{abstract}
Having previously identified the photon field with a (special)
linear Complex, we give a brief account on identifications and reasoning
so far. Then, in order to include spinorial degrees of freedom into 
the Lagrangean description, we discuss the mapping of lines to spins
based on an old transfer principle by Lie. This introduces quaternionic
reps and relates to our original group-based approach by SU(4) and 
SU*(4) $\cong$ SL(2,H), respectively. Finally, we discuss some related
geometrical aspects in terms of (spatial) projective geometry which
point to a projective construction scheme and algebraic geometry.
\end{abstract}


\maketitle


\section{Introduction}
So far, we've transformed our original, group-based view (see 
e.g.~\cite{dahm:aaca} or \cite{dahm:2008}) from using classical
point(like) reps\footnote{As before, we use this shorthand 
notation for 'representation(s)'.} in Lagrangean approaches 
towards a Lagrangean description which includes higher order 
objects as well as 'extended' geometrical objects like lines
and Complexe in order to describe physical observations.

In this context, the major building block has been the identification
of certain geometrical objects, properties and symmetries which 
-- when associated with 'point' reps in $P^{5}$ -- yields line 
geometry in $P^{3}$, i.e.~in real 3-dim space represented by 
homogeneous line coordinates.
We'll summarize few aspects and references in section \ref{ch:DiracTheory}.
Based on this identification, in \cite{dahm:2018a} we have used
\PLu's four line coordinates $(r,\rho,s,\sigma)$ via the (Euclidean)
line rep $x=rz+\rho$, $y=sz+\sigma$ as well as Lie's reasoning 
and transfer principle \cite{lie:1872} between lines and spheres. 
By rearranging line coordinates, we have shown that we thus obtain
a matrix rep in terms of Pauli matrices, i.e.
\begin{equation}
\label{eq:matrixrep}
\left(
\begin{array}{cc} r & \rho\\ s & \sigma\end{array}
\right)
\,\sim\,
\left(
\begin{array}{cc} -Z & X+iY\\ X-iY & +Z\end{array}
\right)
\,\sim\,
X\sigma_{1}-Y\sigma_{2}-Z\sigma_{3}
\end{equation}
relating Lie's two 3-dim spaces $r$ and $R$ (\cite{dahm:2018a}, 
section~III.B). The lhs of eqn.~(\ref{eq:matrixrep}) is based 
on the space $r$ and describes a projective transformation in 
real 3-space with the usual projective geometry, whereas the 
rhs relates to point reps $(X,Y,Z)$ in the space $R$ and Lie's
sphere geometry. We have mapped this sphere geometry to a Pauli
rep (\cite{dahm:2018a}, section~III.B, especially eq.~(7), and 
ibd.~section~III.E), and discussed this 'Lie transfer' and some
related parallels between the respective geometries of $r$ and 
$R$ (\cite{dahm:2018a}, section~III.G ff.). Thus we have related
lines and Complexe with typical SL(2,$\mathbbm{C}$) spinor, or 
quaternionic, calculus. In \cite{dahm:2018a}, we have also shown
that Cartan's spinor calculus has its foundation (and we think 
its origin) in Study's and Beck's work \cite{dahm:2018a}, and 
that these topics have to be treated as a subset of rational 
curves and advanced (projective) geometry (PG).

On the other hand, in \cite{dahm:2018b} -- using the same original
line rep $x=rz+\rho$, $y=sz+\sigma$ -- we have discussed that
Minkowski's paper \cite{mink:1910} on special relativity (SR) 
encapsulates certain aspects of line and projective geometry 
in the contemporarily emerging 4-vector description, and we have
shown that this treatment and invariant theory can be simplified 
by switching to linear Complexe and their geometry. Especially,
the two 'invariants' nowadays derived by the SU(2)$\times$$i$ 
SU(2) interpretation of SR are directly related to the parameters
of a linear Complex \cite{dahm:2018b}.

So here, we want to use those prerequisites to approach Dirac theory 
and various spinor representations commonly used throughout Lagrangean 
descriptions of quantum field theory (QFT). Based on this reasoning,
here we are going to discuss two possible approaches and related 
aspects. In section~\ref{ch:DiracTheory}, we summarize few necessary
aspects of \cite{dahm:2018a} and \cite{dahm:2018b}. In 
section~\ref{ch:diracspinors}, we 're-organize' the spinorial rep 
in a suitable manner for use in sections \ref{ch:alternative} and 
\ref{ch:spin}. In section~\ref{ch:alternative}, we discuss a real 
interpretation of the $\mathbbm{C}_{4\times 2}$ matrix in order to 
visualize the action of the Dirac algebra which can be related to 
the lhs of eq.~(\ref{eq:matrixrep}) and the space $r$. We close in 
section~\ref{ch:spin} with an interpretation related to $R$, and 
a brief outlook.

\section{Group and Rep Theory}
\label{ch:DiracTheory}
In \cite{dahm:MRST4}, appendix 1, we have discussed the situation of
(compact) SU(4) reps where, departing from SO(6), we can connect SU(4)
to various related real forms, and to the Lie algebras so($m$,$n$) 
or groups SO($m$,$n$) with $m+n=6$. This gave a further hint to work
in $P^{5}$ right from the beginning and discuss Complex (and line) 
geometry. So with respect to (non-compact) SU$*$(4) $\cong$ 
SL(2,$\mathbbm{H}$) (covering SO(5,1) twice), or 'Dirac theory' 
and appropriate reps, we want to discuss once more Dirac's original 
problem to find linear reps here: Given the (quadratic) energy-momentum
relation $p_{\mu}p^{\mu}=E^{2}-\vec{p}^{\,2}=m^{2}$, and considering
a rep of $p_{\mu}$ by differential operators acting on a rep $\psi$,
how do we construct\footnote{We restrict the discussion to the 
historical question focusing on the tangent rep and differential
geometry. More general, there are additional possibilities from 
within PG. One approach based on \PLu's line variable $\eta$ will
be given in section~\ref{ch:spin}.} and identify the (linear) rep
spaces and their geometry?

There is (at least) twofold interest in this discussion:

On the one hand, $x_{\mu}x^{\mu}=0$ in point space is related to
a quadratic Complex in line space, $x_{\alpha}$ denoting homogeneous
point coordinates. So given 
$\alpha_{1}x^{2}_{1}+\alpha_{2}x^{2}_{2}+\alpha_{3}x^{2}_{3}+\alpha_{0}x^{2}_{0}=0$
to describe a second order surface\footnote{We suppress a detailed 
discussion of the associated polar theory here!}, the same surface
in line coordinates reads as 
\begin{equation}
\label{eq:quadraticlinecomplex}
\alpha_{1}\alpha_{2}p_{12}^{2}
+\alpha_{1}\alpha_{3}p_{13}^{2}
+\alpha_{2}\alpha_{3}p_{23}^{2}
+\alpha_{0}\alpha_{1}p_{01}^{2}
+\alpha_{0}\alpha_{2}p_{02}^{2}
+\alpha_{0}\alpha_{3}p_{03}^{2}
=0
\end{equation}
In the case of the 'Minkowski metric', $x^{2}_{1}+x^{2}_{2}+x^{2}_{3}-x^{2}_{0}=0$,
eq.~(\ref{eq:quadraticlinecomplex}) yields the quadratic Complex
$p_{12}^{2}+p_{13}^{2}+p_{23}^{2}-p_{01}^{2}-p_{02}^{2}-p_{03}^{2}=0$,
or formally an SO(3,3) symmetry\footnote{We have discussed already
possible complexifications, or 'transfers', to SO($n$,$m$) with 
$n+m=6$ \cite{dahm:MRST4}. The covering groups (have to) change 
appropriately, too.} in $P^{5}$. Note for later use, that this 
description of the surface in terms of line coordinates is defined
only up to a multiple of the \PLL condition 
$P=p_{01}p_{23}+p_{02}p_{31}+p_{03}p_{12}=0$, i.e.~we are allowed 
to add $m P$ to the lhs of eq.~(\ref{eq:quadraticlinecomplex}),
$m$ being a number.

In \cite{dahm:2018a}, section~III.I, we have drawn attention already
to Clebsch's paper \cite{clebsch:1869} on Complex symbolism to 
treat powers of Complexe and associate linear Complexe which may
serve to obtain linear reps. This, however, shifts the focus to 
the (linear) reps themselves\footnote{See also \cite{dahm:MRST4}, 
appendix, with respect to a discussion of reps and rep dimensions
in SU(4) context and a geometrical identification.} as $x_{\mu}$
or $p_{\mu}$ are of dimension 4 while Complexe are of dimension 6.
Whereas we have already associated the photon to a special (linear)
Complex \cite{dahm:2014}, \cite{dahm:2018b} which has to be treated
carefully due to degeneracies and thus appears in different r\^{o}les
and contexts, a regular linear Complex $\textfrak{R}$ is associated 
to a (non-degenerate) correlation, the null system, mapping point to 
plane coordinates, $u_{\beta}=(\textfrak{R})_{\beta\alpha}x_{\alpha}$,
and vice versa. This is automatically related to an incidence relation
$x\cdot u=0$, because 
$x_{\beta}u_{\beta}=x_{\beta}(\textfrak{R})_{\beta\alpha}x_{\alpha}
=(\textfrak{R})_{\beta\alpha}x_{\beta}x_{\alpha}=0$ due to the
antisymmetry of the $\textfrak{R}$ rep\footnote{We want to discuss 
details of this correlation in the comprehensive part VI of this 
series which we expect to be published soon. Right here, it is 
obvious that we may switch to class view instead of working with
orders (and points), so that 
$u_{\alpha}u_{\alpha}
=(\textfrak{R})_{\alpha\gamma}x_{\gamma}(\textfrak{R})_{\alpha\delta}x_{\delta}
=x_{\gamma}(\textfrak{R}^{T})_{\gamma\alpha}(\textfrak{R})_{\alpha\delta}x_{\delta}$
which according to the antisymmetry of $\textfrak{R}$ comprises
not only the 'classical' null systems and forces, but Lie (algebra)
theory as well. On the other hand, we have to relate this to second
order surfaces and their polar theory, so that Clifford algebras
and Dirac's approach emerge as well whereas a representation in
terms of line reps or Complexe has to include generating lines
and appropriate involutions in the tangent plane(s) \cite{clebsch:1891}.}.
While the calculus of SR treats '4-dim points' (and tensorial 
reps thereof), and requires additional assumptions and rules with 
respect to a certain ('Minkowski') metric, quadratic Complexe
are embedded in the well-defined and well-known framework of 
(advanced) projective geometry, so e.g.~the metric can be simply
derived by the Cayley-Klein mechanism. We know from the early 
days from mechanics (e.g.~in M{\"o}bius' work or see \PLL 
\cite{pluecker:1866}), that the central notion of 'a force'
has to be represented by 6-dim line reps and null systems, and
we know from Minkowski and Poincar\'{e} that one should map 
the 6-dim line rep (or more general a linear Complex) to an 
antisymmetric twofold tensor $F^{\mu\nu}$ to incorporate forces
into the 4-dim rep theory of special relativity \cite{mink:1910}
\cite{dahm:2018b}. So why not just go back to the original force
definition\footnote{For example, the $r^{-2}$ dependence of forces
with respect to the (Euclidean) radius can be mapped to quadratic
plane coordinates using the (dual) class picture instead of the
description by orders, i.e.~we use tangent planes to envelop 
the sphere. The radius $r$ may then be written as 
$r=(ua+vb+wc+1)\sqrt{u^{2}+v^{2}+w^{2}}^{-1}$ (\cite{clebsch:1891}, 
p.~26/27), and the sphere with origin in its center reads as 
$r^{-2}=u^{2}+v^{2}+w^{2}$, $u, v, w$ describing Euclidean/affine
plane coordinates. So instead of writing differential expressions
of a potential $\sim r^{-1}$,
we may use simple global expressions as well, of course being 
quadratic in both sets of coordinates, but with global (and not
only infinitesimal) validity.} and see how to extract correct 
reps or even irreps? 

\begin{figure}[h]
\begin{center}
\includegraphics[scale=0.5]{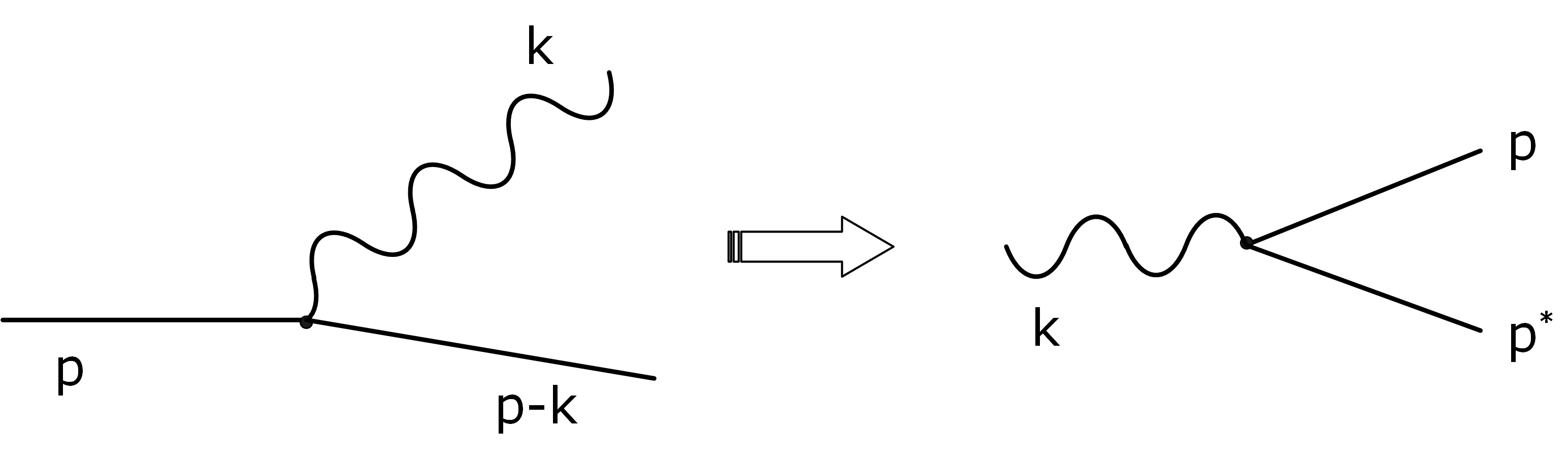}
\end{center}
\caption{\label{fig:gaugevertex}Figure caption}
\end{figure}

On the other hand, the notion of tangents (and as such of lines!) 
has been transported to metric (tangent) spaces, differential 
geometry and 4-dim (point) reps of a 'momentum' $p^{\mu}$, and
used throughout literature, whereas some problematic parts of 
this identification like mass (see e.g.~\cite{sexlurbantke:1992},
ch.~4.1) or the separation of spin (and thus, of course, orbital
angular momentum) from the common 6-dim rep are treated separately
in terms of additional linear reps or (in some approaches) as
'perturbations'. In order to keep both pictures (lines and 4-vector
calculus) alive, we focus on the 6-dim rep, either in terms of 
linear Complexes using line coordinates, or in terms of the twofold
antisymmetric tensor rep $F^{\mu\nu}$ (as in the case of a special
linear Complex in electromagnetism). Even from the viewpoint of 
'classical' gauge theories, this can be pursued using a usual gauge
boson vertex on a (fundamental) spinor, which we might rearrange 
into a gauge boson with momentum $k$ splitting into two conjugate
or even adjoint spinor reps like sketched in Fig.~\ref{fig:gaugevertex}.
Formally, this rises once more the question to find two reps (square 
roots or conjugate objects) which combine to an appropriate boson 
(or vectorial) rep. The major clue, however, will be given in 
section~\ref{ch:spin} when invoking the mighty transfer principles
from classical projective geometry, here mainly Lie transfer, which
in the background may be related to Poncelet's classification of
second order surfaces and their relation to the absolute plane.

\section{Dirac Theory, Spinors and the Point Picture}
\label{ch:diracspinors}
Instead of recalling the historical discussion on how to obtain 
a linear equation of motion\footnote{For details, we refer to 
\cite{bjoedrell} or \cite{lurie:1968}, which serve as our 
references in the text, too.} and appropriate reps, we want to 
separate some hard structural requirements from a plenty of 
technical details which lead to algebraic and analytic 'equations'
being nowadays put in the center of attention, although being 
formally nothing but an artefact of the chosen rep. A simple example
is the discussion of the 'linear' Dirac equation, where the
4-dim momentum rep\footnote{In the upcoming, comprehensive part VI,
we are going to discuss some aspects to identify $p$ in general with
a \protect{\emph{planar}} rep related to polar theory and respecting
duality, and not as derived from point rep generalizations.} $p^{\mu}$
is 'obtained' by acting typically with partial derivatives on 'plane 
waves', $\partial_{\mu}\exp(-ip\cdot x)\sim
\frac{\partial}{\partial x^{\mu}}\exp(-ip\cdot x)\sim
-ip_{\mu}\exp(-ip\cdot x)$ which formally yields the identification
$-i\partial_{\mu}\sim p_{\mu}$ ('quantization'\footnote{We have 
discussed in \cite{dahm:MRST4}, appendix, the dimensionality 4 of
this rep already.}).
So the usual ingredients of this approach are a plane wave borrowed
from physical and Fourier arguments, the requirement of linearity 
applied to the reps to preserve the formal apparatus, and a 
polar-like/tangential reasoning when acting with partial derivatives
on forms in {\it homogeneous} coordinates\footnote{We'll discuss 
some aspects of these requirements later as a special case of the
identification of $p\cdot x$ as well as their intrinsic assumptions.
Instead of equations of motions and (linear) rep theory, it is better
to start from conic sections (or more general a quadric) of energy
and momentum, and discuss linear reps of appropriate square roots.
Or in terms of PG, given a quadric in a plane or in 3-space, how do
we generate the quadric linearly? This 'problem', however, is not 
exhaustive with respect to generation of only the quadric as well 
as to the restriction to possible {\it linear} reps and objects.}.
The problems are shifted into the 'god-given' metric of SR which
corresponds to the related polar form, thus serves to map
$-i\partial_{\mu}\sim p_{\mu}$ by (polar) incidence $x^{\mu}p_{\mu}$=0,
and allows for the (euclidean) identification of 'mass' in the
'momentum'\footnote{Note, that according to the identification
by 6-vectors (or linear Complexe) \cite{mink:1910} \cite{dahm:2018b},
one irrep comprises linear and angular momentum, i.e.~the polar 
{\it and} the axial 3-vector, when decomposed according to affine
and euclidean coordinates.}.

The spinorial rep itself is usually treated in QFT as 'a spinor'
in $\mathbbm{C}_{4\times 1}$ (see e.g.~\cite{bjoedrell}, eq.~(3.2),
or ibd.,~eq.~(3.16), or \cite{lurie:1968}, eq.~1(47)) to fulfil the
equations of motion, although it is well-known that the Dirac 
matrices $(\vec{\alpha},\beta)$ or $(\vec{\gamma},\beta)$ exhibit
their (also well-known) interesting $2\times 2$-block structure. 
As for our purposes (and especially from the viewpoint of 6-dim 
line reps or Complexe as the underlying rep of forces, i.e.~comprising
translational and rotational momenta as well), it is obvious to 
find unified descriptions of (3-dim) momentum and (3-dim) 'spin'
\cite{dahm:2018a}. So we use the 'decomposition' $\mathbbm{C}_{4\times 1}
\longleftrightarrow\mathbbm{C}_{4\times 2}\mathbbm{C}_{2\times 1}$ 
in order to respect the usual block structure of the Dirac matrices
and separate the 'spin' interpretation as discussed in \cite{dahm:2018a},
the more since in practical calculations and formalisms, spin often
is averaged and 'disappears'. So it is usually sufficient to respect
related invariants. Moreover, a {\it very} classical but obviously
long forgotten explanation of 'spin' will be given in section~\ref{ch:spin}.\\

So the next step is to interpret the $\mathbbm{C}_{4\times 2}$
part above on which the Dirac matrices act by their $2\times 2$ block
structure. The direct approach would be to grasp e.g.~the Dirac
spinor reps being indirectly visible in \cite{bjoedrell}, eq.~(3.7),
and the grouping in \cite{bjoedrell}, eq.~(3.16), or directly the 
reps given in \cite{lurie:1968}, eq.~1(51a), 1(51b), in terms of 
objects
\begin{equation}
\label{eq:uv-spinors}
u_{\vec{p}\sigma}=\sqrt{\tfrac{E+m}{2m}}
\left(\begin{array}{c}
\mathbbm{1}\\
\frac{\vec{\sigma}\cdot\vec{p}}{E+m}
\end{array}\right)\xi_{\sigma}
\,,\quad
v_{\vec{p}\sigma}=\sqrt{\tfrac{E+m}{2m}}
\left(\begin{array}{c}
\frac{\vec{\sigma}\cdot\vec{p}}{E+m}\\
\mathbbm{1}
\end{array}\right)\xi_{\sigma}\,,
\end{equation}
$\xi_{\sigma}$ being an appropriately chosen rep in $\mathbbm{C}_{2\times 1}$.
Now, if we act with Dirac/Clifford operators like $\gamma^{i}$, 
$\gamma^{0}$ or $\gamma_{5}$ {\it on the blocks} of this 
structure, the Pauli/quaternion algebra acts within the blocks
and as such on the respective individual block content. Whereas
we want to discuss this block structure by a (real) toy model 
in section~\ref{ch:alternative} and based on previous work 
\cite{dahm:2018a} in section~\ref{ch:spin}, it is the major 
aspect of section~\ref{ch:spin} to reconnect some geometrical 
aspects discussed in \cite{dahm:2018a} to the reps given in 
eq.~(\ref{eq:uv-spinors}). Here, we want to emphasize once
more that in eq.~(\ref{eq:uv-spinors}) the 'mass' obviously
can be divided out, i.e.~we are left with the 4-velocity\footnote{Here,
we use the spinorial notation $u$ and $v$ as well as the 4-velocity
notation $u$ in their respective contexts only to respect common
notation, see e.g.~\cite{sexlurbantke:1992}, ch.~4.1. Our own
notation uses $u$ and $v$ typically as plane coordinates.} 
$u_{\mu}$, only, in the rep.

The interpretation of this rep so far is typically related 
to assumptions and interpretations of negative energy and 
'anti'-particles \cite{bjoedrell}, whereas practically the
$v$-contributions of eq.~(\ref{eq:uv-spinors}) often are 
simply neglected in calculations. More 'extended versions'
transport additional symmetries like chirality or helicity
even to spontaneously broken symmetries and nonlinear reps
\cite{alfaro:1973}, or mix them with compact group structures.
Moreover, the usual treatment includes metric arguments 
and metric interpretations of the variables, combined with 
appropriate 'rules' and 'gauge equations', although the 
'light cone' is known to be an absolute element.

Here, we do not want to extend this criticism, but instead, 
we depart from the reps in eq.~(\ref{eq:uv-spinors}), and
try to gain some more insight (hopefully) independent of 
'the known' physical motivation. So now the simple question
is: Given objects like in eq.~(\ref{eq:uv-spinors}), what
are the possibilities to identify those objects geometrically,
and are these identifications unique? Both answers are simple:
There is definitely more than one well-known possibility, and
as such: NO, of course, it is NOT unique!\\

\section{Real Approach: Line-Coordinates and Transformations}
\label{ch:alternative}
Another important fitting piece of the puzzle can be treated by 
a real toy model. So before proceeding to some applications of 
Lie transfer, this picture can be attached to a special interpretation
of the Dirac 4$\times$2 'spinor' in terms of two 4-dim (point)
'vector' reps, i.e.~we represent 3-dim (real) points in terms
of their four homogeneous coordinates $x_{\alpha}$ and $y_{\beta}$,
respectively, and use the 'Dirac spinor' notation to identify 
the $\left(\begin{smallmatrix}4\\2 \end{smallmatrix}\right)=6$ 
\PLL line coordinates 
$p_{\alpha\beta}=x_{\alpha}y_{\beta}-x_{\beta}y_{\alpha}$.
This can be interpreted as well as identifying the six independent 
2$\times$2 subdeterminants in this 'spinor' rep, and transformations
by the Clifford/Dirac algebra transform the six underlying, basic 
line coordinates of a $P^{5}$. So the 4$\times$2 notation serves as
a 'container' to denote either two (homogeneous) point reps as well
as the related line rep.

\subsection{Some Algebra}
Here, it is of course necessary to investigate the action of Dirac's
gamma matrices on this 4$\times$2 'spinor' and its intrinsic line 
coordinates\footnote{We comment on the Dirac picture, including 'spin'
and 'momentum', as well as on the (quaternionic) Weyl picture in 
section \ref{ch:spin}. Note already here, that using 
the null system as correlation, $x_{\alpha}$ and $y_{\alpha}$ in 
eq.~(\ref{eq:rktwospinor}) can be transferred to planar coordinates
$u_{\alpha}$ and $v_{\alpha}$ as well while preserving the line
interpretation. So with respect to $\psi$, we find an equivalent 
rep $\Psi$ in terms of $u_{\alpha}$ and $v_{\alpha}$. In both
cases, we are left with reps $(\textfrak{R})_{\beta\alpha}$ from 
section~\ref{ch:DiracTheory}. The complete description, however,
has to take care of symmetric (polar) transformations besides the 
null system because $R^{2}=1$ also allows for symmetric reps. So
the diagonal (anti-commutator) part has to be considered besides 
the skew (commutator) part, i.e.~we expect both aspects in the
(Clifford) algebraic description when acting on $u$ and $v$.}
in detail! So defining 
\begin{equation}
\label{eq:rktwospinor}
\psi:=\left(
\begin{array}{cc}
x_{0} & y_{0} \\
x_{1} & y_{1} \\
x_{2} & y_{2} \\
x_{3} & y_{3}
\end{array}
\right)\,,
\end{equation}
and using the reps (see \cite{bjoedrell}) of $\gamma^{\mu}$, $\sigma^{\mu\nu}$
and $\gamma^{5}=\gamma_{5}$, we can identify the following transformed line 
coordinates $p'_{\alpha\beta}\sim A\,p_{\alpha\beta}$, $p'_{\alpha\beta}$
being extracted from the transformed spinor $\psi'$ given in the top row 
of each table\footnote{In the tables, we've arranged the order of the line 
coordinates $p'_{\alpha\beta}$ in appropriate pairs already to simplify and
shorten the subsequent discussion.}, 
\ifTwoCol
\begin{equation}
\label{eq:tablepart1}
\begin{array}{c||c||c|c|c|c|}
p'_{\alpha\beta} & \mathbbm{1}\,\psi & \gamma^{1}\,\psi & \gamma^{2}\,\psi 
& \gamma^{3}\,\psi & \gamma^{0}\,\psi \\
\hline
p'_{01} & +p_{01} & -p_{23} & -p_{23} & -p_{23} & +p_{01} \\
p'_{23} & +p_{23} & -p_{01} & -p_{01} & -p_{01} & +p_{23} \\
p'_{02} & +p_{02} & +p_{13} & -p_{13} & +p_{02} & -p_{02} \\
p'_{13} & +p_{13} & +p_{02} & -p_{02} & +p_{13} & -p_{13} \\
p'_{03} & +p_{03} & +p_{03} & +p_{03} & -p_{12} & -p_{03} \\
p'_{12} & +p_{12} & +p_{12} & +p_{12} & -p_{03} & -p_{12}
\end{array}\,,\\
\end{equation}

\begin{equation}
\label{eq:tablepart2}
\begin{array}{c||c||c|c|c|c|}
p'_{\alpha\beta} & \gamma_{5}\,\psi 
& \gamma_{5}\gamma^{1}\,\psi & \gamma_{5}\gamma^{2}\,\psi 
& \gamma_{5}\gamma^{3}\,\psi & \gamma_{5}\gamma^{0}\,\psi \\
\hline
p'_{01} & +p_{23} 
& -p_{01} & -p_{01} & -p_{01} & +p_{23} \\
p'_{23} & +p_{01}
& -p_{23} & -p_{23} & -p_{23} & +p_{01} \\
p'_{02} & -p_{02}
& -p_{13} & +p_{13} & -p_{02} & +p_{02} \\
p'_{13} & -p_{13}
& -p_{02} & +p_{02} & -p_{13} & +p_{13} \\
p'_{03} & -p_{12}
& -p_{12} & -p_{12} & +p_{03} & +p_{12} \\
p'_{12} & -p_{03}
& -p_{03} & -p_{03} & +p_{12} & +p_{03}
\end{array}\,,\\
\end{equation}

\begin{equation}
\label{eq:tablepart3}
\begin{array}{c||c|c|c|c|c|c|}
p'_{\alpha\beta} 
& \sigma^{01}\,\psi & \sigma^{02}\,\psi & \sigma^{03}\,\psi
& \sigma^{12}\,\psi & \sigma^{13}\,\psi & \sigma^{23}\,\psi \\
\hline
p'_{01} & +p_{23} & +p_{23} & +p_{23} & -p_{01} & -p_{01} & -p_{01} \\
p'_{23} & +p_{01} & +p_{01} & +p_{01} & -p_{23} & -p_{23} & -p_{23} \\
p'_{02} & +p_{13} & -p_{13} & +p_{02} & +p_{02} & -p_{13} & +p_{13} \\
p'_{13} & +p_{02} & -p_{02} & +p_{13} & +p_{13} & -p_{02} & +p_{02} \\
p'_{03} & +p_{03} & +p_{03} & -p_{12} & -p_{03} & +p_{12} & +p_{12} \\
p'_{12} & +p_{12} & +p_{12} & -p_{03} & -p_{12} & +p_{03} & +p_{03}
\end{array}
\end{equation}
\else
\begin{equation}
\label{eq:tablepart1}
\begin{array}{c||c||c|c|c|c||c||c|c|c|c|}
p'_{\alpha\beta} & \mathbbm{1}\,\psi & \gamma^{1}\,\psi & \gamma^{2}\,\psi 
& \gamma^{3}\,\psi & \gamma^{0}\,\psi & \gamma_{5}\,\psi 
& \gamma_{5}\gamma^{1}\,\psi & \gamma_{5}\gamma^{2}\,\psi 
& \gamma_{5}\gamma^{3}\,\psi & \gamma_{5}\gamma^{0}\,\psi \\
\hline
p'_{01} & +p_{01} & -p_{23} & -p_{23} & -p_{23} & +p_{01} & +p_{23} 
& -p_{01} & -p_{01} & -p_{01} & +p_{23} \\
p'_{23} & +p_{23} & -p_{01} & -p_{01} & -p_{01} & +p_{23} & +p_{01}
& -p_{23} & -p_{23} & -p_{23} & +p_{01} \\
p'_{02} & +p_{02} & +p_{13} & -p_{13} & +p_{02} & -p_{02} & -p_{02}
& -p_{13} & +p_{13} & -p_{02} & +p_{02} \\
p'_{13} & +p_{13} & +p_{02} & -p_{02} & +p_{13} & -p_{13} & -p_{13}
& -p_{02} & +p_{02} & -p_{13} & +p_{13} \\
p'_{03} & +p_{03} & +p_{03} & +p_{03} & -p_{12} & -p_{03} & -p_{12}
& -p_{12} & -p_{12} & +p_{03} & +p_{12} \\
p'_{12} & +p_{12} & +p_{12} & +p_{12} & -p_{03} & -p_{12} & -p_{03}
& -p_{03} & -p_{03} & +p_{12} & +p_{03}
\end{array}\quad\text{and}\\
\end{equation}
\begin{equation}
\label{eq:tablepart2}
\begin{array}{c||c||c|c|c|c|c|c|}
p'_{\alpha\beta} & \mathbbm{1}\,\psi
& \sigma^{01}\,\psi & \sigma^{02}\,\psi & \sigma^{03}\,\psi
& \sigma^{12}\,\psi & \sigma^{13}\,\psi & \sigma^{23}\,\psi \\
\hline
p'_{01} & +p_{01} & +p_{23} & +p_{23} & +p_{23} & -p_{01} & -p_{01} & -p_{01} \\
p'_{23} & +p_{23} & +p_{01} & +p_{01} & +p_{01} & -p_{23} & -p_{23} & -p_{23} \\
p'_{02} & +p_{02} & +p_{13} & -p_{13} & +p_{02} & +p_{02} & -p_{13} & +p_{13} \\
p'_{13} & +p_{13} & +p_{02} & -p_{02} & +p_{13} & +p_{13} & -p_{02} & +p_{02} \\
p'_{03} & +p_{03} & +p_{03} & +p_{03} & -p_{12} & -p_{03} & +p_{12} & +p_{12} \\
p'_{12} & +p_{12} & +p_{12} & +p_{12} & -p_{03} & -p_{12} & +p_{03} & +p_{03}
\end{array}
\end{equation}
\fi
This table shows some remarkable features when acting with $\gamma$-matrices
(or 'the Dirac/Clifford algebra') on 4$\times$2 'spinors' $\psi$ of 
eq.~(\ref{eq:rktwospinor}):
\begin{itemize}
\item[-] The six line coordinates (or 2$\times$2 determinants) 
$p_{\alpha\beta}$ are defined by their respective row positions within 
the 4$\times$2 'spinor' only, i.e.~by combinatorics and antisymmetry
from the point (or plane) reps of the underlying coordinate system.
\item[-] Thus the action of the Dirac or Clifford algebra is independent
of the (physical/ma\-the\-ma\-ti\-cal) identification or content of 
the respective variables and coordinates at the respective position 
in the 'spinor', but it depends only on the (row) position and the 
antisymmetry of the line coordinate (or the determinant), and it 
preserves the line coordinate structure within the 4$\times$2 
'spinor'.
\item[-] The transformations of the Dirac algebra rep given by 
\cite{bjoedrell} -- although intrinsically complex -- map real
line coordinates $p_{\alpha\beta}$ to real line coordinates 
$p'_{\alpha\beta}$, independent of the respective content at 
the position in the 4$\times$2 'spinor'. As such, although 
transforming six (real) line coordinates into six (real) line
coordinates, Dirac's approach performs the transformations 
within a 4$\times$4 matrix rep (NOT by means of a 6$\times$6
matrix rep!) which increases the intricacy of the rep but (in 
some special reps) is suitable to represent 4-dim transformations.
\item[-] So in the background, using the 'two-point' interpretation
of the 4$\times$2 'spinor' given above, we perform nothing but
real PG in terms of \PLL coordinates, i.e.~a projective geometry
of 3-space using lines as basic geometric elements in $P^{3}$.
And as such, the generalization leads to Complex geometry in 
$P^{5}$, or equivalently, to the Lie transferred sphere geometry.
\item[-] In other words, the 4$\times$2 'spinor' may serve as a rep
to perform line calculus when acting with linear combinations of
individual base elements of the Dirac algebra on $\psi$, or as a rep
of a linear (line) Complex when acting with linear combinations of
individual base elements of the Dirac algebra on $\psi$ and equating
this action to 0.
\item[-] The 15-dim transformations using real parameters with Dirac
or Clifford algebra elements thus describe nothing but projective
transformations of lines onto lines in 3-dim space, however, one has 
to take care because lines are mapped to lines by duality, and thus 
correlations as well as collineations are described by 15-dim 
transformation groups each, whose algebraic/purely formal effects 
'overlap'.
\item[-] It is necessary to understand and see that the apparent
complexifications in the matrix definitions of the Dirac algebra
elements -- emerging in pairs or $2\times 2$-blocks -- as well as
overall '$i$'s are absorbed in signs by the definition of \PLL 
coordinates in terms of $2\times 2$-determinants of the $4\times 2$
'spinor'. This shifts the question on the meaning of such complexifications
back to point reps and especially to properties of the respective
transfer principle applied in general to the elements of projective
3-space; it cannot be answered from the viewpoint of (2nd order) 
line or polar geometry only.
\item[-] From above, it is obvious that Dirac's approach in terms of
$\gamma$-matrices acting on $\psi$ represents a symbolic scheme (or 
calculus, in the sense of the old German 'Kalk\"{u}l'), and as such 
the trace mechanism and the various relations/formulae between traces
and determinants are justified from above.
\end{itemize}

\subsection{Transformed Views of the Dirac Algebra}
Using the transformation results $p'_{\alpha\beta}$ of the tables
given in eqns.~(\ref{eq:tablepart1}) and (\ref{eq:tablepart2}), 
it is obvious that we can identify the individual coordinate
'positions' of the $p'_{\alpha\beta}$ in a $P^{5}$ rep. As such,
if we 're-group' the coordinates $p'_{\alpha\beta}$ into 'new' 
structures according to $(p_{01},p_{23})$, $(p_{02},p_{13})$, 
and $(p_{03},p_{12})$, then by acting with the Dirac algebra, we 
only find changes in the overall signs and, in some cases, an 
exchange of both line coordinates within these doublets. In order
to quantify the analytic properties, it is thus self-evident 
to introduce the notation\footnote{Note, that with respect to 
the bi-linear invariant, we are free to introduce the additional
notion of adjoint (i.e.~of transposed, conjugated or hermitean)
elements as well, see e.g.~Hamermesh (\cite{hamermesh:1962}, 
p.~369/370), or $\textfrak{A}^{I}=\textfrak{A}^{Adj}\gamma^{0}$,
etc.}
\begin{eqnarray}
\textfrak{A}_{\pm}=(p_{01},\pm p_{23})\,,\quad
\textfrak{B}_{\pm}=(p_{02},\pm p_{13})\,,\quad
\textfrak{C}_{\pm}=(p_{03},\pm p_{12})\\
\textfrak{A}^{I}_{\pm}=(p_{23},\pm p_{01})\,,\quad
\textfrak{B}^{I}_{\pm}=(p_{13},\pm p_{02})\,,\quad
\textfrak{C}^{I}_{\pm}=(p_{12},\pm p_{03})\,.
\end{eqnarray}
Accordingly, the tables in eqns.~(\ref{eq:tablepart1}) and 
(\ref{eq:tablepart2}) can be rewritten as
\begin{equation}
\label{eq:congruencepart1}
\begin{array}{c||c||c|c|c|c|}
& \mathbbm{1}\,\psi & \gamma^{1}\,\psi & \gamma^{2}\,\psi 
& \gamma^{3}\,\psi & \gamma^{0}\,\psi \\
\hline
\textfrak{A}'_{\pm} 
& +\textfrak{A}_{\pm} 
& \mp\textfrak{A}^{I}_{\pm} & \mp\textfrak{A}^{I}_{\pm} & \mp\textfrak{A}^{I}_{\pm}
& +\textfrak{A}_{\pm} \\
\textfrak{B}'_{\pm} 
& +\textfrak{B}_{\pm} 
& \pm\textfrak{B}^{I}_{\pm} & \mp\textfrak{B}^{I}_{\pm} & +\textfrak{B}_{\pm}
& -\textfrak{B}_{\pm} \\
\textfrak{C}'_{\pm} 
& +\textfrak{C}_{\pm} 
& +\textfrak{C}_{\pm} & +\textfrak{C}_{\pm} & \mp\textfrak{C}^{I}_{\pm}
& -\textfrak{C}_{\pm}
\end{array}\,,\\
\end{equation}

\begin{equation}
\label{eq:congruencepart2}
\begin{array}{c||c||c|c|c|c|}
 & \gamma_{5}\,\psi 
& \gamma_{5}\gamma^{1}\,\psi & \gamma_{5}\gamma^{2}\,\psi 
& \gamma_{5}\gamma^{3}\,\psi & \gamma_{5}\gamma^{0}\,\psi \\
\hline
\textfrak{A}'_{\pm} 
& \pm\textfrak{A}^{I}_{\pm} 
& -\textfrak{A}_{\pm} & -\textfrak{A}_{\pm} & -\textfrak{A}_{\pm}
& \pm\textfrak{A}^{I}_{\pm} \\
\textfrak{B}'_{\pm} 
& -\textfrak{B}_{\pm} 
& \mp\textfrak{B}^{I}_{\pm} & \pm\textfrak{B}^{I}_{\pm} & -\textfrak{B}_{\pm}
& +\textfrak{B}_{\pm} \\
\textfrak{C}'_{\pm} 
& \mp\textfrak{C}^{I}_{\pm} 
& \mp\textfrak{C}^{I}_{\pm} & \mp\textfrak{C}^{I}_{\pm} & +\textfrak{C}_{\pm}
& \pm\textfrak{C}^{I}_{\pm}
\end{array}\,,\\
\end{equation}

\begin{equation}
\label{eq:congruencepart3}
\begin{array}{c||c|c|c|c|c|c|}
& \sigma^{01}\,\psi & \sigma^{02}\,\psi & \sigma^{03}\,\psi
& \sigma^{12}\,\psi & \sigma^{13}\,\psi & \sigma^{23}\,\psi \\
\hline
\textfrak{A}'_{\pm} 
& \pm\textfrak{A}^{I}_{\pm} & \pm\textfrak{A}^{I}_{\pm} & \pm\textfrak{A}^{I}_{\pm}
& -\textfrak{A}_{\pm} & -\textfrak{A}_{\pm} & -\textfrak{A}_{\pm} \\
\textfrak{B}'_{\pm} 
& \pm\textfrak{B}^{I}_{\pm} & \mp\textfrak{B}^{I}_{\pm} & +\textfrak{B}_{\pm}
& +\textfrak{B}_{\pm} & \mp\textfrak{B}^{I}_{\pm} & \pm\textfrak{B}^{I}_{\pm} \\
\textfrak{C}'_{\pm} 
& +\textfrak{C}_{\pm} & +\textfrak{C}_{\pm} & \mp\textfrak{C}^{I}_{\pm}
& -\textfrak{C}_{\pm} & \pm\textfrak{C}^{I}_{\pm} & \pm\textfrak{C}^{I}_{\pm}
\end{array}
\end{equation}
Geometrically, the line coordinate sets $\textfrak{A}_{\pm}$, 
$\textfrak{B}_{\pm}$ and $\textfrak{C}_{\pm}$ describe opposite, 
non-intersecting edges of the fundamental tetrahedron in 3-space
when we interpret the individual six line coordinates as base 
'vectors', or basic lines. As such, in the non-degenerate case,
each of the three sets comprises two non-intersecting lines, 
and -- besides polar theory -- we can apply the notion of 
Congruences\footnote{We do not want to blur the discussion, or 
be imprecise or vague, however, out of various possibilities 
to identify the geometrical setup, here, we discuss only two 
identifications: Two opposite edges of the fundamental tetrahedron,
each can be seen as axis of a special Complex, so lines fulfilling 
both constraints satisfy a Congruence (see also \cite{dahm:2014}). 
The second case, in section~\ref{sec:relaspects}, is devoted 
to a linear Complex in the additive rep, i.e.~forming one 
constraint in line coordinates. Related, there is deep historic
background, see e.g.~\cite{kleinHG:1926}.} \cite{plueckerNG:1868}. 
So even the interchange of the line coordinates 
$p_{\alpha\beta}\longleftrightarrow p_{\gamma\delta}$, 
$\alpha\beta\neq\gamma\delta$ (or 
$p_{\alpha\beta}\sim\epsilon_{\alpha\beta\gamma\delta}p_{\gamma\delta}$)
within the sets according to 
$\textfrak{A}_{\pm}\longrightarrow\textfrak{A}^{I}_{\pm}$, 
$\textfrak{B}_{\pm}\longrightarrow\textfrak{B}^{I}_{\pm}$ and 
$\textfrak{C}_{\pm}\longrightarrow\textfrak{C}^{I}_{\pm}$ 
exchanges only the two lines within each of the Congruences
but respects the Congruence itself and doesn't mix it with 
the two other Congruences. This is valid for the full action
of the Dirac/Clifford algebra. In this context, if we stress 
the notion of irreps, with respect to the Dirac algebra, we 
have thus found an irreducible structure in terms of three 
individual Congruences, and it depends on the (line) coordinates
(or the linear Complex) to specify the Congruence in detail.
It is noteworthy, that in the language of QFT, we have found
threefold intrinsic symmetry structures in terms of the three
'independent' Congruences above. However, the interpretation 
and the explanation is purely geometric in terms of projective
geometry while using lines and Complexe as base elements.

\subsection{Related Aspects}
\label{sec:relaspects}
If we represent the Complexe in 'classical (additive) form', 
i.e.~$\textfrak{A}=a p_{01}+b p_{23}$, we can represent the 
action of the transformations above as well. Some of the 
transformations leave the line coordinates invariant, some
change the signs pairwise, so $p_{01}\longrightarrow\pm p_{01}$
and $p_{23}\longrightarrow\pm p_{23}$ which yields 
$\textfrak{A}=\pm (a p_{01} + b p_{23})=\pm\textfrak{A}$.
So these transformations change the variables $a\longrightarrow\pm a$
and $b\longrightarrow\pm b$ which in effect may only change
the sign of the Complex $\textfrak{A}$.

Some transformations interchange the line coordinates 
$p_{01}\longleftrightarrow p_{23}$ as discussed above which 
in fact leads to an inversion of the parameters,
i.e.~$(\frac{a}{b},1)\longrightarrow(\frac{b}{a},1)$,
or with $\lambda = \frac{a}{b}$, we have 
$(\lambda,1)\longrightarrow(\lambda^{-1},1)$. So on the 
line we find reciprocity $\lambda\longleftrightarrow\lambda^{-1}$.
In general, we see in (\ref{eq:congruencepart1}) -- 
(\ref{eq:congruencepart3}), that, if we 'absorb' the changes 
$\textfrak{A}\longleftrightarrow\textfrak{A}^{I}$
(and similarly with $\textfrak{B}$, and $\textfrak{C}$) in 
an inversion of the respective coefficient(s) $\lambda$, then 
$\textfrak{A}_{\pm}\longrightarrow\textfrak{A}_{\pm}$,
$\textfrak{B}_{\pm}\longrightarrow\textfrak{B}_{\pm}$,
and $\textfrak{C}_{\pm}\longrightarrow\textfrak{C}_{\pm}$. 
The action of the Dirac/Clifford algebra doesn't change or 
mix $\textfrak{A}$, $\textfrak{B}$, and $\textfrak{C}$, and
moreover, it respects and doesn't change the subscripts $\pm$.
Thus we have found another justification of Klein's six fundamental
Complexe and SU(2)$\times$SU(2) symmetry with respect to their 
3$\oplus$3 handedness. So dependent on the original rep $\psi$, 
the transformation of the Dirac algebra respects a threefold 
substructure induced by the geometrical properties of the 
fundamental (coordinate) tetrahedron when expressed in line 
coordinates.

With respect to further physical applications, there is however an 
additional and very interesting aspect\footnote{We want to thank 
B. Schmeikal for helpful private discussions and remembering Onsager
theory.} which we may thus connect to Complexe and SU(4). Here, we
discuss the thesis \cite{elchaar} as a bridge to related topics and
further applications of algebra theory in physics, and we cannot go
into details. However, \cite{elchaar}, chapter~4.1 yields a definition
of a Lie algebra with (skew) generators\footnote{We have adapted 
to our notation.} $X_{\alpha\beta}$ denoted as 'tetrahedron algebra'.
After having defined the 'three-point $sl_{2}$ Loop algebra'
and stated a Lie algebra isomorphism, a related homomorphism
$\psi$ has been given/cited \cite{elchaar} by
\begin{equation}
\begin{array}{lcl}
\psi(X_{12})=x\otimes 1 & \, & \psi(X_{03})=y\otimes t  +z\otimes(t  -1)\\
\psi(X_{23})=y\otimes 1 & \, & \psi(X_{01})=z\otimes t' +x\otimes(t' -1)\\
\psi(X_{31})=z\otimes 1 & \, & \psi(X_{02})=x\otimes t''+y\otimes(t''-1)
\end{array}
\end{equation}
with $t'=1-t^{-1}$, $t''=(1-t)^{-1}$, and $x$, $y$, $z$ from
$sl_{2}$ ('equitable basis'). The important point from our
discussion above, however, is the identification of their operators
$u_{0}$, $u_{1}$, and $u_{2}$ by $4u_{0}=\psi(X_{02}+X_{31})$, 
$4u_{1}=\psi(X_{03}+X_{12})$, and $4u_{2}=\psi(X_{01}+X_{23})$
as generators of $sl_{2}\otimes k[t,t^{-1},(1-t)^{-1}]$ being 
a Lie algebra over $k$. The labels of every two non-adjacent 
edges generate a subalgebra isomorphic to the Onsager algebra,
i.e.~(\cite{elchaar}, Prop.~4.2.4) for mutually distinct
$\alpha, \beta, \gamma, \delta \in [0,1,2,3]$, the subalgebra
of the 'tetrahedron algebra' generated by $X_{\alpha\beta}$ 
and $X_{\gamma\delta}$ is isomorphic to the Onsager algebra.
The 'tetrahedron algebra' thus is a (direct) sum of three
Onsager algebras.

By comparison, the definitions of $X_{\alpha\beta}$ and their 
figure~4.1 (\cite{elchaar}, p.~77), the operators $X$ obviously
represent classical line coordinates algebraically. Understanding
the homomorphism $\psi$ as transfer, $u_{0}$, $u_{1}$, and $u_{2}$
correspond to appropriate reps of our Congruences above, i.e.~from
our point of view \cite{elchaar} yields another transfer of 
projective geometry to a special notion.

\section{Spin}
\label{ch:spin}
Now with respect to general transformations parametrized by real numbers
in the Dirac algebra, already here it is obvious that we have switched
to a coordinate description using lines as base elements (respectively 
the appropriate \PLL coordinates). So by going back to the well-known
Dirac algebra description, with respect to the real toy model of section
\ref{ch:alternative} there are two open issues at this time: We have 
used two points to define our 'spinor' $\psi$ in terms of their
homogeneous coordinates, so how does the rep given e.g.~in \cite{bjoedrell}
-- as well as the various other spinorial reps floating around in
literature -- relate to this description or fit into this picture?
Moreover, thinking in terms of (4-dim) quaternions (and considering
their conjugates as well), we can also define $4\times 2$ spinors 
while switching to their complex $2\times 2$ rep, e.g.~in terms of
complexified Pauli matrices, and use SL(2,$\mathbbm{H}$), or SU$*$(4)
rep theory.\\

For now, however, we want to depart once more from\footnote{For us,
to connect to PG above, it seems more natural to use the surface
$\vec{p}^{\,2}-E^{2}$, as can be seen from the minus sign of $\eta$
in the next footnote.} $p_{\mu}p^{\mu}=E^{2}-\vec{p}^{\,2}=m^{2}$,
or $u_{\mu}u^{\mu}=1$ using 4-velocities.

If we recall from \cite{dahm:2018a}, section~III.E, our equations~(16)
and (17),
\begin{equation}
\label{eq:inhomogvars}
r=\frac{1}{2}(\pm H'-Z')\,,
\rho=\frac{1}{2}(X'+iY')\,,
s=\frac{1}{2}(X'-iY')\,,
\sigma=\frac{1}{2}(\pm H'+Z')
\end{equation}
and
\begin{equation}
\label{eq:maptopauli}
\frac{1}{2}\left(
\begin{array}{cc} -Z'\pm H' & X'+iY'\\ X'-iY' & +Z'\pm H'\end{array}
\right)
\sim
X'\sigma_{1}-Y'\sigma_{2}-Z'\sigma_{3}\pm H'\sigma_{0}
\,,
\end{equation}
we have shown there, that \PLu's fifth coordinate yields
$\eta=r\sigma-s\rho=H'^{2}-Z'^{2}-X'^{2}-Y'^{2}$.
The variables $X'$, $Y'$, $Z'$ were point coordinates of Lie's 
second, transferred 3-dim space $R$, $H'$ denoted a 'sphere radius'
and plays a special r\^{o}le which we are going to discuss elsewhere
in more detail. For analytic calculations and reps, we can use the
geometry in $R$-space (or 'the spin geometry') which we began to 
develop in \cite{dahm:2018a}, III.G ff. 

So if we now choose to define $\eta=p_{\mu}p^{\mu}$ (although being
in $R$-space but relying on Lie's transfer), and $\eta$ being the 
determinant of the line coordinates in $r$-space, then $\eta$
maps to a squared mass, i.e.~$\eta=m^{2}$. If instead we use 
unimodular reps, $\eta=1$, then $\eta=u_{\mu}u^{\mu}$ yields
4-velocities, and a quadratic equation as well.

In both approaches, $\eta$ is quadratic (which \PLL needed to 
preserve the grade of his coordinates during transformations),
and we may (formally) introduce a 'spinorial picture' by two
'doublets' $(r,s)$ and $(\rho,\sigma)$ according to
\begin{equation}
\label{eq:linesinglet}
\eta=
\left(\begin{array}{cc}
r & s 
\end{array}\right)
\left(\begin{array}{cc}
0 & +1\\ 
-1&0
\end{array}\right)
\left(\begin{array}{c}
\rho\\
\sigma
\end{array}\right)\,,\quad
\left(\begin{array}{cc}
0 & +1\\ 
-1&0
\end{array}\right)
\sim i\,,\quad i^{2}=-1\,,
\end{equation}
so that $\eta$ seems to behave as a 'singlet' (which corresponds
to an interpretation of the complex structure $i$ above). However,
care has to be taken in that $r,s,\rho$ and $\sigma$ are
inhomogeneous line coordinates in $r$-space, and the full formal
treatment within PG\footnote{If we 'translate back', the 'radius'
$H'=r+\sigma$ \cite{dahm:2018a}, III.E, reads as 
$H=(p_{01}+p_{23})p_{03}^{-1}$, and the value -- being a ratio 
-- is fixed and (usually) finite. So $H'=0$ denotes the Complex
$p_{01}+p_{23}=0$ in $r$-space (i.e.~in real 3-dim space) which
we already know from above. $\eta=-p_{12}p_{03}^{-1}$, so this 
also amounts to a finite and measurable value which we have to
compare to mass calculations. Note the $-$-sign in $\eta$!} has
to use (homogeneous) line coordinates $p_{\alpha\beta}$.

For now, in the context of finding linear reps with respect to 
expressions like $p_{\mu}p^{\mu}$ it should be noted that line 
geometry yields such a possibility by eqn.~(\ref{eq:linesinglet}).
The origin of such an approach, of course, is located in the 
theory of second order surfaces and their generation by lines.

Last not least, if we rewrite $\eta=m^{2}\mathbbm{1}=
E^{2}\mathbbm{1}-(\vec{\sigma}\cdot\vec{p}\,)^{2}$, we are 
formally\footnote{In order to perform a direct comparison
to Dirac and the spinor reps (\ref{eq:uv-spinors}), recall
that we want to resolve $E^{2}$ into linear reps.} free to
discuss a 'massless' case $\eta=0$ as well as finite $\eta$.
Then $E^{2}\mathbbm{1}=(\vec{\sigma}\cdot\vec{p}\,)^{2}
+m^{2}\mathbbm{1}$, and we can cast this expression (even 
if we change $m^{2}\longrightarrow -m^{2}$) in various forms.
Common is the decomposition of the quadratic term into two
components $(a\mathbbm{1}+\vec{\sigma}\cdot\vec{p}\,)$ and
$(a\mathbbm{1}-\vec{\sigma}\cdot\vec{p}\,)$ which we may 
arrange in a two-component 'spinor' form, and thus mix them
also by matrix transformations into other (linear) reps. In 
case of
$\eta=m^{2}\mathbbm{1}=
E^{2}\mathbbm{1}-(\vec{\sigma}\cdot\vec{p}\,)^{2}$, we obtain
$a^{2}=(E^{2}-m^{2})\mathbbm{1}$, and the equation reads 
$a^{2}\mathbbm{1}-(\vec{\sigma}\cdot\vec{p}\,)^{2}=0$.
The square roots of $\eta=m^{2}$ are real quaternions,
$E^{2}\mathbbm{1}-(\vec{\sigma}\cdot\vec{p}\,)^{2}=
E^{2}\mathbbm{1}+(-i\vec{\sigma}\cdot\vec{p}\,)^{2}=
E^{2}\mathbbm{1}+(\vec{q}\cdot\vec{p}\,)^{2}$, so we recover
SL(2,$\mathbbm{H}$) rep theory. The square roots of $E^{2}$
are the bi-quaternions known from QFT, as 
$E^{2}=m^{2}+(\vec{\sigma}\cdot\vec{p}\,)^{2}
=m^{2}-(-i\vec{\sigma}\cdot\vec{p}\,)^{2}
=m^{2}-(\vec{q}\cdot\vec{p}\,)^{2}
=(m+\vec{q}\cdot\vec{p}\,)(m-\vec{q}\cdot\vec{p}\,)
=m^{2}+(i\vec{q}\cdot\vec{p}\,)^{2}$.\\

From the last section, it is obvious that phases (and '$i$'s)
matter. So considering various notations in QFT, from above,
we have no freedom to determine phases freely. As we have 
pointed out in the footnote, a 'natural' choice of the $\eta$
would include a $-$-sign, or an additional $i^{2}$. So there 
is work left to compare to the various different reps around.
However, it is apparent that the origin of Dirac's approach
can be located in classical PG of second order (and class) 
surfaces, and that line (and Complex) geometry is the general
framework which we want to enhance in upcoming work.

\end{document}